# Identification of Key Proteins Involved in Axon Guidance Related Disorders: A Systems Biology Approach

Ishtiaque Ahammad[1]*

1. Department of Biochemistry and Microbiology, North South University, Dhaka, Bangladesh.
* Corresponding author. Email: **ishtiaque.ahammad@northsouth.edu**

**Abstract**

Axon guidance is a crucial process for growth of the central and peripheral nervous systems. In this study, 3 axon guidance related disorders, namely- Duane Retraction Syndrome (DRS), Horizontal Gaze Palsy with Progressive Scoliosis (HGPPS) and Congenital fibrosis of the extraocular muscles type 3 (CFEOM3) were studied using various Systems Biology tools to identify the genes and proteins involved with them to get a better idea about the underlying molecular mechanisms including the regulatory mechanisms. Based on the analyses carried out, 7 significant modules have been identified from the PPI network. Five pathways/processes have been found to be significantly associated with DRS, HGPPS and CFEOM3 associated genes. From the PPI network, 3 have been identified as hub proteins- DRD2, UBC and CUL3.

**Introduction**

Duane Retraction Syndrome (DRS) is a human genetic disorder of cranial nerve guidance (Engle 2010). It's a disorder of ocular motility involving deficient horizontal eye movements, eyelid retraction, palpebral fissure narrowing in adduction and a variety of other abnormal movement of the affected eye when the other eye is fixated in various cardinal positions (SHALABY and BAHGAT 2010; ALEXANDRAKIS and SAUNDERS 2001) . The absence of $6^{th}$ cranial nerve and the consequent developmental adaptation that occurs in the embryo results in the abnormal pattern of ocular motility of DRS (SHALABY and BAHGAT 2010).

Horizontal Gaze Palsy with Progressive Scoliosis (HGPPS) is a clinically and genetically homogeneous disorder which is associated with failure of hindbrain axons to cross the midline

(Engle 2010). Symptoms include being born with restricted horizontal gaze and development of scoliosis within the first decade of life. It results from mutations in the ROBO3 gene and is autosomal recessive in nature (Jen et al. 2004).

Congenital fibrosis of the extraocular muscles type 3 (CFEOM3), an autosomal dominant congenital eye movement disorder involves variable unilateral or bilateral ophthalmoplegia, limited vertical ductions, and blepharoptosis (drooping eyelids) (Mackey et al. 2002).

The purpose of this study was to identify crucial genes and proteins associated with Duane Retraction Syndrome (DRS), Horizontal Gaze Palsy with Progressive Scoliosis (HGPPS) and Congenital fibrosis of the extraocular muscles type 3 (CFEOM3) to get a better idea about the molecular mechanism, pathways and the key players. In order to do this, DRS, HGPPS and CFEOM3 associated genes from several databases were retrieved. Various Systems Biology tools have been put to use to carry out several analyses such as pathway enrichment analysis, protein-protein interaction (PPI) network, module analysis and identification of transcriptional regulators.

## Methods

### DRS, HGPPS and CFEOM3 associated genes

The keywords "Duane Retraction Syndrome", "HGPS", and "Congenital fibrosis of the extraocular muscles type 3" were used to screen out the genes associated with them from 3 databases and the results were put together to build a set of DRS, HGPPS and CFEOM3 associated genes. The three databases used for finding the genes were-

- GeneCards (version 3.0), a searchable, integrative database which provides comprehensive information on all annotated and predicted human genes (Safran et al. 2010).
- Search Tool for Interacting Chemicals (STICH, version 5.0), a database of known and predicted interactions between chemicals and proteins (Szklarczyk et al. 2016)
- Comparative Toxicogenomics Database (CTD), which provides manually curated information concerning chemical–gene/protein interactions, chemical–disease as well as gene–disease relationships was used (Davis et al. 2017).

### PPI Network

Data from 2 interaction databases were merged in order to predict the Protein-Protein Interaction (PPI) pairs among the DRS, HGPPS and CFEOM3 associated genes. The databases were-

- Biological General Repository for Interaction Datasets (BioGRID, version 3.4, https://wiki.thebiogrid.org/) (Chatr-aryamontri et al. 2015)
- The Molecular Interaction Database (MINT, 2012 update, https://mint.bio.uniroma2.it/) (Licata et al. 2012)

Using the software Cytoscape (http://www.cytoscape.org) (Saito et al. 2012), a PPI network was visualized for DRS, HGPPS and CFEOM3 associated genes.

Using the CytoNCA plug-in (Li et al. 2017) (version 2.1.6, http://apps.cytoscape.org/apps/cytonca) in Cytoscape, degree centrality (DC), betweenness centrality (BC), and closeness centrality of the nodes of the PPI network were subjected to analysis to identify the hub proteins (He and Zhang n.d.). "Without weight." was set as the parameter.

**Module Analysis**

For module analysis of the PPI network, MCODE plug-in (Bader and Hogue 2003) (version 1.4.2; http://apps.cytoscape.org/apps/mcode; parameters set as degree cut-off = 2, maximum depth = 100, node score cut-off = 0.2, and -core = 2) in Cytoscape was used.

**Pathway Enrichment Analysis**

KEGG pathway enrichment analysis for the nodes of top modules was carried out with JEPETTO plug-in (Winterhalter, Widera, and Krasnogor 2014) in Cytoscape.

**Identification of Transcriptional Regulators**

Transcription factors (TFs) of DRS, HGPPS and CFEOM3 associated genes were searched and then their targets were identified using the transcriptional regulatory relationships unravelled by a sentence-based text-mining (TRRUST, http://www.grnpedia.org/trrust/) (Han et al. 2015) database.

## Results and Discussion

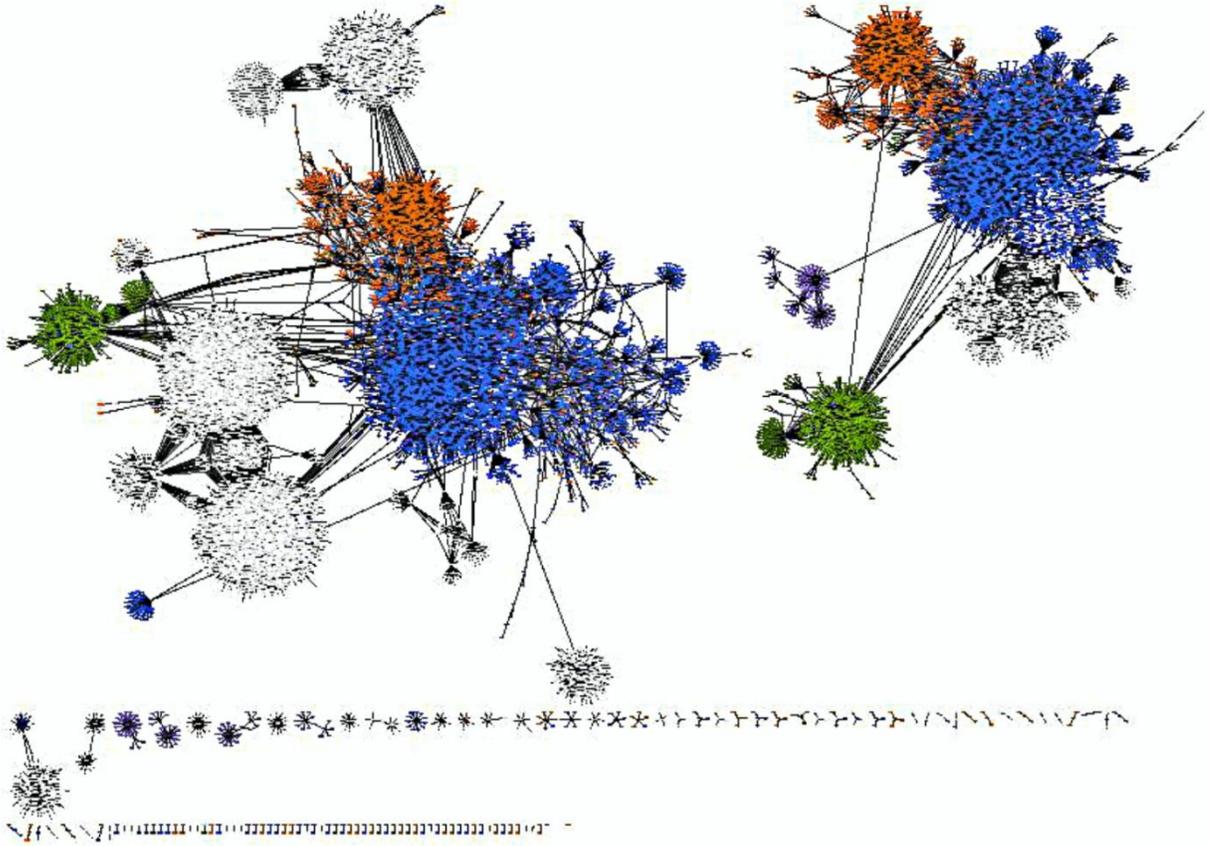

**Figure 1: Protein-protein interaction network of with DRS, HGPPS and CFEOM3 associated genes. It contains 24,062 nodes (proteins) and 29,226 edges (interactions).**

**Table 1: Axon guidance related disorder associated genes**

| Duane Retraction Syndrome (DRS) | Horizontal Gaze Palsy with Progressive Scoliosis (HGPPS) | Congenital fibrosis of the extraocular muscle type3 (CFEOM3) |
|---|---|---|
| CHN1, MAFB, SALL4, SALL1, PAX2, TBX5, CHD7, KIF21A, BMP4, TBX3, TUBB3, ADNP, DPM1, HOXA1, SALL2, CUL3, DURS1, CA8, FOXL2, COL25A1, CPA6, FGF9, PHOX2A, CECR2, CDX2, IFNG, ARX, CREBBP, KIR2DL1, HLA-DRB5, SCD5, MOCS3, BCAS4, KIR2DL4, KIR3DL3, CHRNB3, REV3L, PLXND1, PIEZO2, MBS1, MBS3, MT-TA, EYA1, FGF13, EP300, | ROBO3, PHOX2A, KIF21A, UBC CHL1, ELOC, ELOB, CHL1-AS2, ROBO3 | TIMP2, TGFBI |
| L1CAM, DRD2, GJA1, NEUROG1, EPHA4, STAT1, KIR3DP1, HOXD3, GZMB, UBA2, ECEL1, ADAM17, PTPRN2, ROBO3, TOLLIP, XIRP2, BTRC, LBX1, HGS, HOXD1 | | |

**Table 2: Identification of Hub proteins**

| Degree Centrality | | Betweenness Centrality | | Closeness Centrality | |
|---|---|---|---|---|---|
| Gene | Score | Gene | Score | Gene | Score |
| DRD2 | 3560.0 | DRD2 | 1.3852736E8 | EGFR | 1.2288813E-4 |
| DRD2 | 3226.0 | DRD2 | 9.7794936E7 | ARRB2 | 1.2288793E-4 |
| UBC | 1609.0 | UBC | 6.9540432E7 | UBC | 1.2288745E-4 |
| ADAM17 | 1273.0 | ADAM17 | 4.7671964E7 | DRD2 | 1.2288734E-4 |
| PHAZE | 1215.0 | CUL3 | 4.2251952E7.0 | CALM2 | 1.2288698E-4 |
| UBC | 1188.0 | EGFR | 3.4354112E7 | CALM2 | 1.2288698E-4 |
| CUL3 | 1176.0 | UBC | 2.9233554E7 | CALM3 | 1.2288698E-4 |
| Ubc | 817.0 | UBC | 2.7864122E7 | CUL3 | 1.2288621E-4 |
| AI194771 | 785.0 | UBC | 2.567804E7 | FLINA | 1.2288503E-4 |
| UBC | 719.0 | ARRB2 | 2.0969964E7 | HGS | 1.2288491E-4 |

**Table 3: Module analysis**

| Module 1 | Module 2 | Module 3 | Module 4 | Module 5 | Module 6 | Module 7 |
|---|---|---|---|---|---|---|
| ASB-3, ASB7, C2orf5, CCT-epsilon, CCT2, CCT3, CCTETA, COPS6, CUL4A, Cctq, EWS, GFPT2, HSAL2, KIF21A, MRXHF2, NCOA-62, OABP, OCT4, P1.1-MCM3, PHA2E, POP1, RP11-435J19.1, SIII, SRB, TCP20, p59ILK | ARRB2, GRB2, HGS, NEDD8, TRAF1, UBC | DmelCG12081, CG9291, CG4204, l(3)neo31 | Bor, BB138287, Neurog1 | CUL3, DAPK1, Kelch-like ECT2-interacting protein, PIG53, Protein pp110, TSG101, UBC | CBP, Early E1A 32 kDa protein, KAT2B | SIII p15, SIII p18, Ywhae |

**Table 4 : Enrichment results**

| Pathway or Process | XD-score | Q-value | Overlap/size |
|---|---|---|---|
| Dorso-ventral axis formation | 0.18515 | 1.00000 | 1/20 |
| Phototransduction | 0.16697 | 1.00000 | 1/22 |
| Olfactory transduction | 0.14515 | 1.00000 | 1/25 |
| Bladder cancer | 0.09042 | 1.00000 | 1/38 |
| Nucleotide excision repair | 0.08039 | 1.00000 | 1/42 |
| Notch signaling pathway | 0.08039 | 1.00000 | 1/42 |
| Endometrial cancer | 0.06515 | 1.00000 | 1/50 |
| Non-small cell lung cancer | 0.06358 | 1.00000 | 1/51 |
| Acute myeloid leukemia | 0.06208 | 1.00000 | 1/52 |
| Endocytosis | 0.05257 | 1.00000 | 3/178 |

**Table 5: Key transcription factors**

| # | Transcription Factor | Description | Target Genes | Mode of Regulation |
|---|---|---|---|---|
| 1 | CIITA | class II, major histocompatibility complex, transactivator | CREBBP | Unknown |
| | | | HLA-DRB5 | Unknown |
| | | | STAT1 | Unknown |
| 2 | TFAP4 | transcription factor AP-4 (activating enhancer binding protein 4) | IFNG | Unknown |
| | | | SALL2 | Activation |
| 3 | RFX5 | regulatory factor X, 5 (influences HLA class II expression) | HLA-DRB5 | Unknown |
| | | | IFNG | Unknown |
| 4 | SOX2 | SRY (sex determining region Y)-box 2 | BMP4 | Repression |
| | | | TUBB3 | Repression |
| 5 | POU5F1 | POU class 5 homeobox 1 | BMP4 | Repression |
| | | | CDX2 | Repression |
| 6 | REST | RE1-silencing transcription factor | STAT1 | Repression |
| | | | TUBB3 | Repression |
| | | | TUBB3 | Unknown |
| 7 | GATA3 | GATA binding protein 3 | CDX2 | Activation |
| | | | IFNG | Unknown |
| 8 | STAT3 | signal transducer and activator of transcription 3 (acute-phase response factor) | IFNG | Repression |
| | | | SALL4 | Activation |
| | | | STAT1 | Activation |

| | | | GJA1 | Activation |
|---|---|---|---|---|
| 9 | JUN | jun proto-oncogene | IFNG | Activation |
| | | | IFNG | Unknown |
| | | | TIMP2 | Unknown |
| 10 | EP300 | E1A binding protein p300 | CREBBP | Unknown |
| | | | IFNG | Activation |

In this study, a total of 76 DRS, HGPPS and CFEOM3 associated genes were identified from GeneCards, STICH and CTD databases (Table 1) and their protein-protein interaction network was constructed (Figure 1). Based on Betweenness Centrality, Closeness Centrality, and Degree Centrality scores DRD2, UBC and CUL3 were established as hub nodes in the Protein-Protein Interaction network of these genes and their interactions (Table 2). Seven distinct modules (Module 1, Module 2, Module 3, Module 4, Module 5, Module 6 and Module7) of the PPI network were identified (Figure 2). Dorso-ventral axis formation, Phototransduction, Olfactory transduction, Bladder cancer and Nucleotide excision repair were the top 5 pathways/processes the proteins in these modules were found to be significantly involved in (Table 3). Top 10 transcription factors targeting DRS, HGPPS and CFEOM3 associated genes have been identified from TRRUST database of transcription factors (Table 5).

One of the key players identified in this study is the D2 dopamine receptor (DRD2). It is one of the most extensively studied genes implicated in neuropsychiatric disorders. Since its association with the TaqI A DRD2 minor (A1) allele with severe alcoholism in 1990, a number of studies have been carried out (Noble 2003). DR2 have been found to be associated with schizophrenia (Shaikh et al. 1994), posttraumatic stress disorder (PTSD) (Comings et al. 1991, 1996), movement disorders (Planté-Bordeneuve et al. 1997) and so on. UBC is responsible for maintaining cellular ubiquitin levels under stress conditions (Ryu et al. 2007). Transcription of UBC gene is induced during times of stress and helps removing damaged/unfolded proteins (Ryu et al. 2007). Disruption of ubiquitination a mechanism which controls many aspects of neuronal function by regulating protein abundance in neurons has been demonstrated in neurological disorders such as Parkinson's disease, Amyotrophic Lateral Sclerosis and Angleman Syndrome (Hallengren, Chen, and Wilson 2013). CUL3 has been found to be a key regulators of sleep homeostasis and a dopamine arousal pathway in drosophila (Pfeiffenberger and Allada 2012). CUL3 is a member of the Cullin family of proteins whose job is to function as scaffold proteins of E3 ubiquitin ligase complexes. One study showed revealed cell-autonomous involvement of CUL3 in axonal arborization and dendritic elaboration of Drosophila mushroom body neurons. CUL3 mutant neurons were found to be defective in terminal morphogenesis of neurites (Zhu et al. 2005).

Therefore, the three key proteins identified in this study had been implicated in various neurological functions and disorders before but their role in axon guidance related disorders were not elucidated. This new result from the perspective of Systems Biology will hopefully inspire *in vitro* investigators to extract further information about the role of these proteins in propagating diseases like DRS, HGPPS and CFEOM3.

**Conclusion**

To conclude, a total of 76 DRS, HGPPS and CFEOM3 associated genes have been identified. DRD2, UBC and CUL3 have been revealed as key players in DRS, HGPPS and CFEOM3 associated pathways.

**References**

ALEXANDRAKIS, G, and R SAUNDERS. 2001. "DUANE RETRACTION SYNDROME." *Ophthalmology Clinics of North America* 14 (3). Elsevier:407–17. https://doi.org/10.1016/S0896-1549(05)70238-8.

Bader, Gary D, and Christopher WV Hogue. 2003. "An Automated Method for Finding Molecular Complexes in Large Protein Interaction Networks." *BMC Bioinformatics* 4 (1). BioMed Central:2. https://doi.org/10.1186/1471-2105-4-2.

Chatr-aryamontri, Andrew, Bobby-Joe Breitkreutz, Rose Oughtred, Lorrie Boucher, Sven Heinicke, Daici Chen, Chris Stark, et al. 2015. "The BioGRID Interaction Database: 2015 Update." *Nucleic Acids Research* 43 (D1). Oxford University Press:D470–78. https://doi.org/10.1093/nar/gku1204.

Comings, D E, B G Comings, D Muhleman, G Dietz, B Shahbahrami, D Tast, E Knell, P Kocsis, R Baumgarten, and B W Kovacs. 1991. "The Dopamine D2 Receptor Locus as a Modifying Gene in Neuropsychiatric Disorders." *JAMA* 266 (13):1793–1800. http://www.ncbi.nlm.nih.gov/pubmed/1832466.

Comings, D E, R J Rosenthal, H R Lesieur, L J Rugle, D Muhleman, C Chiu, G Dietz, and R Gade. 1996. "A Study of the Dopamine D2 Receptor Gene in Pathological Gambling." *Pharmacogenetics* 6 (3):223–34. http://www.ncbi.nlm.nih.gov/pubmed/8807661.

Davis, Allan Peter, Cynthia J. Grondin, Robin J. Johnson, Daniela Sciaky, Benjamin L. King, Roy McMorran, Jolene Wiegers, Thomas C. Wiegers, and Carolyn J. Mattingly. 2017. "The Comparative Toxicogenomics Database: Update 2017." *Nucleic Acids Research* 45 (D1). Oxford University Press:D972–78. https://doi.org/10.1093/nar/gkw838.

Engle, Elizabeth C. 2010. "Human Genetic Disorders of Axon Guidance." *Cold Spring Harbor Perspectives in Biology* 2 (3). Cold Spring Harbor Laboratory Press:a001784. https://doi.org/10.1101/cshperspect.a001784.

Hallengren, Jada, Ping-Chung Chen, and Scott M Wilson. 2013. "Neuronal Ubiquitin Homeostasis." *Cell Biochemistry and Biophysics* 67 (1). NIH Public Access:67–73. https://doi.org/10.1007/s12013-013-9634-4.

Han, Heonjong, Hongseok Shim, Donghyun Shin, Jung Eun Shim, Yunhee Ko, Junha Shin,

Hanhae Kim, et al. 2015. "TRRUST: A Reference Database of Human Transcriptional Regulatory Interactions." *Scientific Reports* 5 (1). Nature Publishing Group:11432. https://doi.org/10.1038/srep11432.

He, Xionglei, and Jianzhi Zhang. n.d. "Why Do Hubs Tend to Be Essential in Protein Networks?" Accessed March 11, 2018. https://doi.org/10.1371/journal.pgen.0020088.

Jen, Joanna C, Wai-Man Chan, Thomas M Bosley, Jijun Wan, Janai R Carr, Udo Rüb, David Shattuck, et al. 2004. "Mutations in a Human ROBO Gene Disrupt Hindbrain Axon Pathway Crossing and Morphogenesis." *Science (New York, N.Y.)* 304 (5676). NIH Public Access:1509–13. https://doi.org/10.1126/science.1096437.

Li, Min, Yu Lu, Zhibei Niu, and Fang-Xiang Wu. 2017. "United Complex Centrality for Identification of Essential Proteins from PPI Networks." *IEEE/ACM Transactions on Computational Biology and Bioinformatics* 14 (2). IEEE Computer Society:370–80. https://doi.org/10.1109/TCBB.2015.2394487.

Licata, Luana, Leonardo Briganti, Daniele Peluso, Livia Perfetto, Marta Iannuccelli, Eugenia Galeota, Francesca Sacco, et al. 2012. "MINT, the Molecular Interaction Database: 2012 Update." *Nucleic Acids Research* 40 (D1):D857–61. https://doi.org/10.1093/nar/gkr930.

Mackey, David A., Wai-Man Chan, Christopher Chan, W. Gillies, Anne M. Brooks, Justin O'Day, and Elizabeth C. Engle. 2002. "Congenital Fibrosis of the Vertically Acting Extraocular Muscles Maps to the FEOM3 Locus." *Human Genetics* 110 (5):510–12. https://doi.org/10.1007/s00439-002-0707-5.

Noble, Ernest P. 2003. "D2 Dopamine Receptor Gene in Psychiatric and Neurologic Disorders and Its Phenotypes." *American Journal of Medical Genetics* 116B (1):103–25. https://doi.org/10.1002/ajmg.b.10005.

Pfeiffenberger, Cory, and Ravi Allada. 2012. "Cul3 and the BTB Adaptor Insomniac Are Key Regulators of Sleep Homeostasis and a Dopamine Arousal Pathway in Drosophila." *PLoS Genetics* 8 (10). Public Library of Science:e1003003. https://doi.org/10.1371/journal.pgen.1003003.

Planté-Bordeneuve, V, D Taussig, F Thomas, G Said, N W Wood, C D Marsden, and A E Harding. 1997. "Evaluation of Four Candidate Genes Encoding Proteins of the Dopamine Pathway in Familial and Sporadic Parkinson's Disease: Evidence for Association of a DRD2 Allele." *Neurology* 48 (6):1589–93. http://www.ncbi.nlm.nih.gov/pubmed/9191771.

Ryu, Kwon-Yul, René Maehr, Catherine A Gilchrist, Michael A Long, Donna M Bouley, Britta Mueller, Hidde L Ploegh, and Ron R Kopito. 2007. "The Mouse Polyubiquitin Gene UbC Is Essential for Fetal Liver Development, Cell-Cycle Progression and Stress Tolerance." *The EMBO Journal* 26 (11). European Molecular Biology Organization:2693–2706. https://doi.org/10.1038/sj.emboj.7601722.

Safran, M., I. Dalah, J. Alexander, N. Rosen, T. Iny Stein, M. Shmoish, N. Nativ, et al. 2010. "GeneCards Version 3: The Human Gene Integrator." *Database* 2010 (0). Oxford University Press:baq020-baq020. https://doi.org/10.1093/database/baq020.

Saito, Rintaro, Michael E Smoot, Keiichiro Ono, Johannes Ruscheinski, Peng-Liang Wang, Samad Lotia, Alexander R Pico, Gary D Bader, and Trey Ideker. 2012. "A Travel Guide to Cytoscape Plugins." *Nature Methods* 9 (11). Nature Publishing Group:1069–76. https://doi.org/10.1038/nmeth.2212.

Shaikh, S, D Collier, M Arranz, D Ball, M Gill, and R Kerwin. 1994. "DRD2 Ser311/Cys311 Polymorphism in Schizophrenia." *Lancet (London, England)* 343 (8904):1045–46. http://www.ncbi.nlm.nih.gov/pubmed/7909081.

SHALABY, KARIMA L., and MOSTAFA BAHGAT. 2010. "Duane Retraction Syndrome." *The Medical Journal of Cairo University* 78 (2). http://www.erepository.cu.edu.eg/index.php/MJCU/article/view/818.

Szklarczyk, Damian, Alberto Santos, Christian von Mering, Lars Juhl Jensen, Peer Bork, and Michael Kuhn. 2016. "STITCH 5: Augmenting Protein–chemical Interaction Networks with Tissue and Affinity Data." *Nucleic Acids Research* 44 (D1). Oxford University Press:D380–84. https://doi.org/10.1093/nar/gkv1277.

Winterhalter, Charles, Paweł Widera, and Natalio Krasnogor. 2014. "JEPETTO: A Cytoscape Plugin for Gene Set Enrichment and Topological Analysis Based on Interaction Networks." *Bioinformatics* 30 (7):1029–30. https://doi.org/10.1093/bioinformatics/btt732.

Zhu, Sijun, Rosanne Perez, Marc Pan, and Tzumin Lee. 2005. "Requirement of Cul3 for Axonal Arborization and Dendritic Elaboration in Drosophila Mushroom Body Neurons." *The Journal of Neuroscience : The Official Journal of the Society for Neuroscience* 25 (16). Society for Neuroscience:4189–97. https://doi.org/10.1523/JNEUROSCI.0149-05.2005.